\documentclass[twocolumn,epjc3]{svjour3}  
\smartqed  
\RequirePackage{graphicx}
\journalname{Eur. Phys. J. C}

\usepackage{color}
\usepackage[makeroom]{cancel}
\usepackage[normalem]{ulem}

\usepackage[utf8]{inputenc}
\usepackage[T1]{fontenc}
\usepackage{array}
\usepackage{bm}
\usepackage{dsfont}
\usepackage{mathtools}
\usepackage{mathrsfs}
\usepackage[squaren,Gray]{SIunits}
\usepackage{array}
\usepackage{booktabs}
\usepackage{widetable}
\usepackage{widetext}
\usepackage{appendix}
\usepackage{bbold}

\usepackage{graphicx}

\definecolor{kkcolor}{rgb}{0,0.6,0.4}

\definecolor{cali_color}{rgb}{0,0.4470,0.7410}

\definecolor{pkcolor}{rgb}{0,0.1,0.7}

\begin{document}

\newcommand{\subsubsubsection}[1]{\paragraph{#1}\mbox{}\\}
\setcounter{secnumdepth}{4}
\newcommand{\stat}{\textrm{stat}\,}
\newcommand{\sech}{\textrm{sech}\,}
\newcommand{\csch}{\textrm{csch}\,}
\newcommand{\kin}{\textrm{kin}\,}
\newcommand{\spin}{\textrm{spin}\,}
\newcommand{\bare}{\textrm{bare}\,}
\newcommand{\pole}{\textrm{pole}\,}
\newcommand{\ren}{\textrm{ren}\,}
\newcommand{\match}{\textrm{match}\,}
\newcommand{\HQET}{\textrm{HQET}\,}
\newcommand{\QCD}{\textrm{QCD}\,}
\newcommand{\msbar}{\overline{MS}\,}

\newcommand{\oo}{\mathcal{O}}
\newcommand{\lat}{\mathrm{lat}}
\newcommand{\free}{\mathrm{free}}
\newcommand{\massless}{\mathrm{massless}}
\newcommand{\cont}{\mathrm{cont}}
\newcommand{\MSb}{\overline{\mathrm{MS}}}
\newcommand{\MSt}{\widetilde{\mathrm{MS}}}

\newcommand{\latt}{\mathrm{latt}}


\title{Finite volume effects in the McLerran-Venugopalan initial condition for the JIMWLK equation}

\author{Piotr Korcyl\thanksref{e1,addr1}}
\thankstext{e1}{e-mail: piotr.korcyl@uj.edu.pl}

\institute{Institute of Theoretical Physics, Jagiellonian University, ul. \L ojasiewicza 11, 30-348 Krak\'ow, Poland \label{addr1}}

\date{Received: date / Accepted: date}

\maketitle

\begin{abstract}
We revisit the numerical construction of the initial condition for the dipole amplitude from the Mc\-Lerran-Venugopalan model in the context of the JIM\-WLK evolution equation. We observe lar\-ge finite volume effects induced by the Poisson equation formulated on a torus. We show that the situation can be partially cured by introducing an infrared regularization. We propose a procedure which has negligible finite volume corrections. The control of the finite volume and finite lattice spacings effects is crucial when considering the numerical solutions of the JIMWLK evolution equation with the collinear improvement.

\end{abstract}

\section{Introduction}

Evolution equations are necessary elements of the theoretical frameworks developed to describe data from Deep Inelastic Scattering (DIS) experiments gathered from various facilities worldwide. Experimental data was collected for a wide range of kinematical variables describing this process, in particular the momentum transfer $Q^2$ and longitudinal momentum fraction $x$. Our understanding of Quantum Chromodynamics, the theory behind these experiments, provides us with predictions on the dependence of the DIS cross-section on $Q^2$ and $x$ and is summarized by the mentioned evolution equations (see Ref.~\cite{Kovchegov:2012mbw} for a review). Due to the complexity of QCD the known evolution equations are based on perturbative kernels and are valid only in a limited region of phase space. In many situations these evolution equations can be cast into the form of parabolic differential equation and introduce a Hamiltonian governing the dynamics in the evolution variable. In that case, solving the problem requires the knowledge of the initial condition at some initial value of $x_0$ or $Q_0$. That object is inherently non-perturbative and therefore usually it is not available and has to be taken from some approximate model. 

In the case of the JIMWLK equation \cite{Balitsky:1995ub,JalilianMarian:1997jx,JalilianMarian:1997gr,JalilianMarian:1997dw,Kovner:2000pt,Kovner:1999bj,Weigert:2000gi,Iancu:2000hn,Ferreiro:2001qy}, which motivated our present work, the initial condition is provided by the parton probability distribution in the transverse momentum. The description of the DIS data requires only the simplest, two-point gluon probability distribution also known as the dipole distribution (which also enters other processes, see for example Ref.~\cite{2019}). Typically, one resorts to the Color Glass Condensate effective theory of QCD (see e.g. \cite{Gelis:2010nm}) which offers a computational basis where the dipole distribution can be modelled for instance in the McLerran-Venugopalan model \cite{McLerran:1993ni,McLerran:1993ka}.

The JIMWLK evolution equation being a non-linear equation does not admit general analytically known solutions. One can employ numerical methods \cite{Rummukainen:2003ns} to study its solutions for a particular initial condition. The first step of such calculation is the numerical construction of the initial dipole amplitude in a discretized setup. In what follows we show the details of that step and expose the problems associated with finite volume effects which were so far overlooked. We then propose a simple solution and present numerical evidence for its effectiveness. 

The construction of the initial condition in the MV model was already discussed in the Literature (see \cite{initial1995,initial1995b,initial1999,initial2000,initial2001,initial2007,PhysRevC.79.024909}), also from the numerical perspective (for example in Ref.~\cite{Fukushima:2007yk,fukushima2008randomness}), however we found that some aspects concerning finite volume effects were missing and we aim to close this gap with this work. In Ref.~\cite{Lappi:2007ku,Lappi:2016qhk} the numerical values of the parameters in the MV model were set in physical units through numerical calculations. The construction of initial condition based on phenomenological considerations were discussed for example in Refs.~\cite{albacete1,albacete2} and many others.

Our findings and observations are particularly important in the context of collinearly improved JIMWLK evolution equation \cite{cl2009,cl2014,cl2015,cl2015b}. It is well-known since many years that the JIMWLK evolution equation alone is not compatible with experimental data and the remedy to that is expected to come from resummation of single and double logarithms in $Q^2$. As was discussed in Ref.~\cite{Hatta:2016ujq}, such resummation can be introduced into the Langevin formulation of the JIMWKL equation with the help of time-ordering of the consecutive gluon interaction during the evolution and some modifications of the JIMWLK kernel. In order to be able to study the effects of these improvements one has to be able to disentangle them from the various effects associated with the initial condition. This would not be possible with the initial dipole amplitude motivated by the McLerran-Venugopalan model derived numerically with the numerical setup commonly used due to large finite volume effects. In contrast, our proposal for the initial condition does not suffer from any spurious finite volume and finite lattice spacing effects. Hence, any signs of such effects can be readily associated to the physics of the JIMWLK equation and the collinear improvement. The discussion of the solutions of the JIMWLK equation with the collinear improvement are beyond the scope of this work and will be presented elsewhere. 

General information about the open-access numerical code used in this work can be found in Ref.~\cite{Korcyl:2020orf} and in the associated \verb[git[ repository. Complementary study of the systematic effects  of the solutions of the JIMWLK equations was presented in Ref.~\cite{Cali:2021tsh}. All calculations in that reference were conducted at a single value of the volume, hence the conclusions presented there do not touch the issues contained in the current work.

\section{Dipol gluon probability distribution}
\label{sec. dipole}

In the McLerran-Venugopalan (MV) model \cite{McLerran:1993ka,McLerran:1993ni,MV3} the color charge density $\rho^a(\mathbf{x})$ is defined on the infinite transverse plane and follows an independent Gaussian distribution at each point,
\begin{equation}
    \langle \rho^a(\mathbf{x}) \rho^b(\mathbf{y}) \rangle =  \delta_{a,b} \  \delta^{(2)}(\mathbf{x} - \mathbf{y}) \ \mu^2,
\end{equation}
where we dropped the dependence on $x^+$ and $x^-$ light-cone coordinates because it will not play any role in the following discussion.
The $\mu$ parameter is the main parameter of the MV model and has the interpretation of the density of color charges per area element. Following the philosophy of the CGC effective theory, the static color charges represent the large-$x$ gluons inside the probed hadron whereas the small-$x$ gluons result from these charges via the Yang-Mills equations. The latter reduce to the Poisson equation,
\begin{equation}
    \nabla^2 A^{a}(\mathbf{x}) =  \rho^a(\mathbf{x}) \,,
\label{eq. poisson continuum}
\end{equation}
which can be solved for the color potential $A^a(\mathbf{x})$ using the known Green function of the two-dimensional Poisson equation for any given $\rho^a(\mathbf{x})$. Since the interplay of boundary conditions with the existence of solutions will be crucial in what follows let us be more explicit.

On the infinite transverse plane (following the original MV construction) there exists a Green function for the two-dimensional Poisson equation,
\begin{equation}
    G(\mathbf{x},\mathbf{z}) = \frac{1}{2\pi} \ln \big| \mathbf{x} - \mathbf{z} \big|.
    \label{eq: green function}
\end{equation}
Therefore, for any color charges configuration we can obtain the color potential,
\begin{equation}
    A^a(\mathbf{x}) = \int_{-\infty}^{\infty} d\mathbf{z} \ \rho^a(\mathbf{z}) \ G(\mathbf{x},\mathbf{z}).
\end{equation}
Proceeding, we can calculate the expectation value of the product of two color potentials,
\begin{multline}
    \langle A^a(\mathbf{x}) A^b(\mathbf{y}) \rangle = \\ = \int_{-\infty}^{\infty} d\mathbf{z} \ d\mathbf{z'} \ G(\mathbf{x},\mathbf{z}) \ G(\mathbf{y},\mathbf{z'}) \ \langle \rho^a(\mathbf{z}) \rho^b(\mathbf{z'}) \rangle = \\ =
    \mu^2 \  \delta_{a,b} \ \Gamma(\mathbf{x},\mathbf{y}),
    \label{eq: A correlator}
\end{multline}
where
\begin{equation}
    \Gamma(\mathbf{x},\mathbf{y}) = \int_{-\infty}^{\infty} d\mathbf{z} \ G(\mathbf{x},\mathbf{z}) \ G(\mathbf{y},\mathbf{z}).
\end{equation}
Because,
\begin{equation}
    \frac{1}{2\pi} \ln \big| \mathbf{x} \big| = \int \frac{d^2\mathbf{k}}{4\pi^2} \frac{e^{i \mathbf{k} \cdot \mathbf{x}}}{k^2},
    \label{eq: gluon propagator}
\end{equation}
which is the free gluon propagator, one can also see Eq.\eqref{eq: A correlator} as a two-gluon interaction of the hadron with the static color potential. Introducing Wilson lines $U(\mathbf{x})$,
\begin{equation}
    U(\mathbf{x}) = e^{i  A^a(\mathbf{x}) \lambda^a},
\end{equation}
one can calculate
\begin{multline}
    C_{\textrm{MV}}(|\mathbf{x} - \mathbf{y}|) = \langle \textrm{Tr} U^{\dagger}(\mathbf{x}) U(\mathbf{y}) \rangle =\\= \exp\Big( - \frac{(\mathbf{x} - \mathbf{y})^2}{2R^2} \ln\big( \sqrt{2} \frac{\kappa R}{|x|} + e\big)\Big),
    \label{eq: S_MV}
\end{multline}
where the dimensionful combination $\kappa R$ corresponds to the infrared regulator and can be thought of being of the order of $\Lambda_{\textrm{QCD}}$. We express it using the distance $R$ which is in fact a function of the MV parameters, $R = R(g^2 \mu)$. $R$ is the most convenient quantity to be used to set the scale. Preferably, $R$ and $\kappa$ should be set by fitting the gluon distribution to some experimental data.

One should note that there exist also a simpler model for the two-point gluon initial distribution, the Golec-Biernat-Wusthoff (GBW)  model \cite{GolecBiernat:1998js} where the following shape of the initial correlation function is postulated,
\begin{multline}
    C_{\textrm{GBW}}(|\mathbf{x} - \mathbf{y}|) = \langle \textrm{Tr} U^{\dagger}(\mathbf{x}) U(\mathbf{y}) \rangle =\\= \exp\Big( - \frac{(\mathbf{x} - \mathbf{y})^2}{2R^2}\Big).
    \label{eq: S_GBW}
\end{multline}

\begin{figure}[h]
    \centering
    \includegraphics[width=0.49\textwidth]{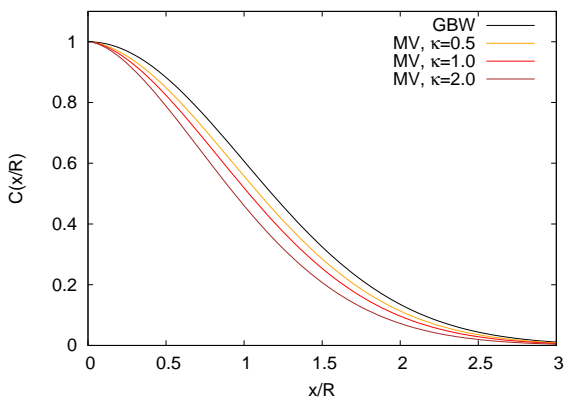}
    \caption{Comparison of the dipole amplitude as a function of the distance for the GBW and MV models.}
    \label{fig: GBW and MV}
\end{figure}

In both cases the dimensionful parameter $R$ has the interpretation of the saturation radius. It is typical to define it for a distribution of any shape as the distance at which the two-point gluon correlator has the value $e^{-\frac{1}{2}}$,
\begin{equation}
    C(|\mathbf{x} - \mathbf{y}| = R_s ) = e^{-\frac{1}{2}}.
    \label{eq: saturation radius}
\end{equation}

We illustrate the differences between $C_{\textrm{GBW}}$ and $C_{\textrm{MV}}$ in Fig.~\ref{fig: GBW and MV}. These two definitions, Eqs.~\eqref{eq: S_MV} and \eqref{eq: S_GBW}, are the usual starting point of analytical calculations of various gluon and quark distributions needed for estimating more cross-sections of process like DIS or with more complex kinematics. If one intends to employ numerical formulation and solve evolution equations numerically, the initial gluon distribution has to be constructed numerically. 

\section{Discretization of the McLerran-Venugopalan model}

In order to treat the problem numerically one has to discretize the transverse plane, i.e. introduce a finite spacing between neighbouring sites. The numerical framework imposes also another fundamental constraint: the calculations are performed in a finite volume. As is typically done, in order to preserve translational symmetry and reduce boundary effects, one applies periodic boundary conditions which transform the infinite transverse plane into a torus. It should be stressed that the finite lattice spacing and the finite volume are \emph{technical} tools which allow to tackle the problem on a computer and hence should not have any effects on the final results, because they do not appear in the original formulation of the models discussed in the previous section. In particular, final results should be obtained after the continuum extrapolation of the lattice spacing to zero and after the infinite volume extrapolation with the radii of the torus going to infinity. In the current problem the two limits are independent and can be taken in any order. Let us finally remark that the results can only be correct if they are properly defined at finite values of the lattice spacing and volume. In order to better illustrate the last statement let us describe the way the MV model is typically discretized. 

One starts with the discretized torus with the color charges $\rho^a(\mathbf{x})$ located at each site. Due to periodicity we have,
\begin{equation}
    \rho^a(\mathbf{x}) \equiv \rho^a(x,y) = \rho^a(x+L_x,y) = \rho^a(x,y+L_y),
\end{equation}
where $L_x$ and $L_y$ are the radii of the torus in physical units, i.e. fm. In order to simplify the discussion we set them equal, $L_x = L_y \equiv L$. The basic object in the calculation are again the Wilson lines
at fixed position $\mathbf{x}$,
$U(\mathbf{x})$. This time $\mathbf{x}$ belongs to a discrete set,
\begin{equation}
    \mathbf{x} = (x,y) = a (n_x, n_y), \qquad n_x, n_y \in \mathbb{Z}
\end{equation}
Wilson line variables $U(\mathbf{x})$ can be obtained from the gauge potentials $A^{a}$
\begin{equation}
    U(\mathbf{x}) = \exp\left( -i g A^{a}(\mathbf{x}) \lambda^a \right) \, ,
\label{eq. exponentiation}
\end{equation}
which in turn are related to the color charges through the Poisson equation, as in Eq.\eqref{eq. poisson continuum}
\begin{equation}
    U(\mathbf{x}) = \exp\left( -i g A^{a}(\mathbf{x}) \lambda^a \right) = 
    \exp\left(-i \frac{g \rho^{a}(\mathbf{x}) \lambda^a}{\nabla^2} \right) \, .
\label{eq. exponentiation2}
\end{equation}
At this point the finite volume and the imposed boundary conditions play a critical role. The finite volume and in particular the fact that the torus is a compact manifold imply that only a total zero charge is allowed by the Gauss law. As a consequence, the Poisson equation does not admit solutions on a torus for a non-zero probe charge, i.e. the Green function Eq.~\eqref{eq: green function} does not exist. This remains true for any value of the lattice spacing, also in the continuum limit. There are two ways of circumventing this problem. 

One can impose the global condition on the random color charges that the total charge is zero. That is physically well motivated as all hadrons are color singlets, and hence it seems reasonable to model them with a charge distribution which has zero net charge. However, this introduces global correlations among different positions in the transverse plane which are not taken into account in the original MV model and are not present in $S_{\textrm{MV}}$ given by Eq.~\eqref{eq: S_MV}. On the technical level this is usually achieved by dropping the zero mode of the Fourier-transformed charge distribution \cite{fukushima2008randomness,Marquet:2016cgx}. Note that this procedure does not suppress long-range gluon interactions and hence can be expected to exhibit large finite volume effects. 

The second way is to regularize the Poisson equation in such a way that would allow solutions for a charge distribution with a non-zero total charge on a compact manifold. Note that our final results correspond to the infinite volume limit where the boundary conditions become irrelevant and the compactness in lost. Hence, the relevant physical correlations should be recovered when the regulator is removed and the infinite volume limit is taken. Obviously, these two limits do not commute, for a given value of the regulator one has to take the infinite volume limit first and only afterwards remove the regulator.

Most of the Literature follows the combination of the two approaches. One can circumvent the problem with finite volume effects by limiting the charge distribution to a small region of the transverse plane. This was investigated in Ref.~\cite{PhysRevD.98.034013} where the size of the proton $B$ was introduced which exponentially suppresses the long range interactions. In that case, when $B$ is much smaller then $L$, finite volume effects are under control.

In this work we keep the Fourier transform unaltered and we introduce the infrared regulator into the Poisson equation. In that case Eq.~\eqref{eq. exponentiation2} becomes
\begin{equation}
    U(\mathbf{x}) = \exp\left( -i g A^{a}(\mathbf{x}) \lambda^a \right) = 
    \exp\left(-i \frac{g \rho^{a}(\mathbf{x}) \lambda^a}{\nabla^2 - m^2} \right) \, .
\label{eq. exponentiation3}
\end{equation}
One can provide a physical interpretation of the infrared regulator parameter $m$ in terms of the gluon propagator. We substitute $1/k^2$ by $1/(k^2+m^2)$ with the aim of introducing by hand the effects of color confinement of gluons at the scale of hadron size. We recover the original MV model when $L\rightarrow \infty$ and $m\rightarrow 0$. We demonstrate how this works in practice in the next Section.

Let us highlight a point that turns out to be important in the following discussion. Namely, that the Green function for the regularized Poisson equation on an infinite plane is known and is proportional to the $K_0$  Bessel function,
\begin{equation}
    G(\mathbf{x},\mathbf{z}, m^2) = \frac{1}{2\pi} K_0 \big( m^2 |\mathbf{x} - \mathbf{z}| \big),
    \label{eq: massive green function}
\end{equation}
which for small $m$ has the asymptotics, $G(\mathbf{x},\mathbf{z}, m^2) \sim \ln \frac{1}{2} m^2 |\mathbf{x} - \mathbf{z}|$. Hence, for sufficiently small values of the infrared regulator we expect to recover the original MV model on the infinite plane.

Let us close this Section by providing last details needed for the numerical implementation of the MV model. Technically, on each site $\mathbf{x}$ we construct a random matrix from the SU(3) algebra by 
\begin{equation}
\rho(\mathbf{x}) = \rho(\mathbf{x})^a \lambda^a \,,
\end{equation}
where the color sources $\rho(\mathbf{x})^a$ are normally distributed and generated by the Box-Muller method. For their standard deviation, we have
\begin{equation}
\langle g \rho(\mathbf{x})^a  g \rho(\mathbf{y})^b \rangle = g^4\mu^2 \delta^{ab}  \delta(\mathbf{x} - \mathbf{y})  \,.
\end{equation}
Above, $g^2$ and $\mu$ are input parameters of the MV model and it turns out that the only relevant dimensionful combination is $g^2 \mu$. All dimensionful quantities can be expressed in units of $g^2 \mu$, in particular the saturation radius $R_s g^2 \mu$ and the circumference of the torus $L g^2 \mu$ become dimensionless numbers. $\mu$ being the value of the charge density parametrizes the line of constant physics, so larger $L g^2 \mu$ corresponds to a larger circumference. Hence, the infinite volume limit is obtained when $L g^2 \mu \rightarrow \infty$. 

\section{Numerical results for the MV model on the torus}

\begin{figure}
    \centering
    \includegraphics[width=0.45\textwidth]{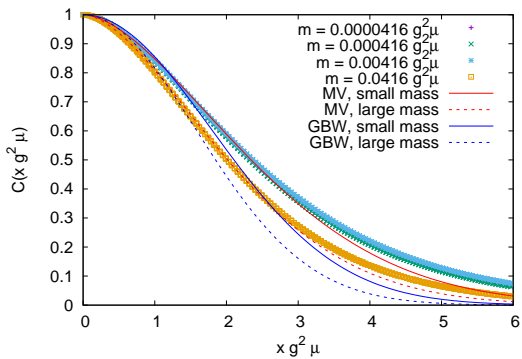}
    \caption{Dipole amplitude generated from the discretized MV model on the torus with an infrared regulator in a small volume $g^2 \mu L = 30.72$. Calculation done using a linear extend of $L/a=512$.}
    \label{fig:MV_initial_condition_small_volume}
\end{figure}

\begin{figure}
    \centering
    \includegraphics[width=0.45\textwidth]{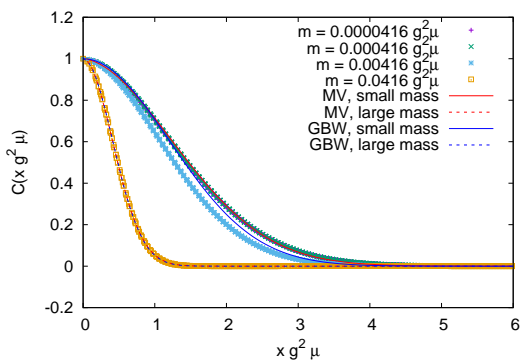}
    \caption{Dipole amplitude generated from the discretized MV model on the torus with an infrared regulator in a large volume $g^2 \mu L = 983.04$. Calculation done using a linear extend of $L/a=16384$.}
    \label{fig:MV_initial_condition_large_volume}
\end{figure}

With the MV model implemented as described above one can calculate the dipole amplitude,
\begin{equation}
    C(|\mathbf{x} - \mathbf{y}|) = \langle \textrm{Tr} U^{\dagger}(\mathbf{x}) U(\mathbf{y}) \rangle
    \label{eq: correlation function}
\end{equation}
and check if the resulting shape corresponds to the predictions of Eq.~\eqref{eq: S_MV}. We present the results in Figs.~\ref{fig:MV_initial_condition_small_volume} and \ref{fig:MV_initial_condition_large_volume} for a wide range of the infrared regulator $m$ values. The former shows the situation in a small volume whereas the latter in a large volume, where small and large corresponds to the value of the $g^2\mu L$ parameter. The same physics should be obtained with the same value of $g^2\mu$ in physical units, hence if $g^2\mu L_1 < g^2\mu L_2$ corresponds to $L_1 < L_2$ in physical units. The typical values used in the Literature vary between $5 \le g^2\mu L \le 200$ \cite{initial2000,initial2001}. For Fig.\ref{fig:MV_initial_condition_small_volume} we set $g^2 \mu L = 30.72$ as was done in recent publications Ref.~\cite{Marquet:2016cgx,Lappi:2014wya,Mantysaari:2018zdd}, whereas in Fig.~\ref{fig:MV_initial_condition_large_volume} we set $g^2 \mu L = 983.04$ which is 32 times larger than in Fig.~\ref{fig:MV_initial_condition_small_volume}. 

Apart of data sets we also show the results of our attempts to identify the shapes predicted by Eqs.~\eqref{eq: S_MV} and \eqref{eq: S_GBW}. We have chosen the lattice spacing small enough that the difference between the continuum extrapolated data is smaller than the statistical uncertainties which in turn are smaller than the symbol size. Hence, if the numerical construction is valid and is performed at the volume large enough one should recover unintegrated gluon distribution of Eq.\eqref{eq: S_MV} for some value of $R$ and $\kappa$. After performing the fit, we see that in the small volume neither of the theoretical predictions describes the data for any of the tested values of the infrared regulator. We note that this parameter was changed by four orders of magnitude and only the largest value exhibits some deviations; other three data sets lie on top of each other.

Fig.~\ref{fig:MV_initial_condition_small_volume} allows us to conclude the for the three smallest values of the regulator we are in the logarithmic regime of the Bessel function and hence the  gluons are not constrained from traveling around the torus. We expect that such long range effects introduce large finite volume effects, especially when the volume is small. Only the data set with the the largest value of the $m$ parameter shows some effect of that regulator. However, for neither of the values of the regulator do we recover the expected shape of the gluon distribution.

Let us now discuss Fig.~\ref{fig:MV_initial_condition_large_volume} where, for the same lattice spacing and the same values of the $m$ parameter as in Fig.~\ref{fig:MV_initial_condition_small_volume}, we show the results from a much larger volume. We notice that the differences between the different values of $m$ are much larger. The two smallest values fall on top of each other, whereas for the larger two the gluon distributions exhibit different shapes. At small values of $m$ the data can be described by the $C_{\textrm{MV}}$ prediction and \emph{cannot} be described by the $C_{\textrm{GBW}}$ formula. We take this as the evidence, that from the point of view of correlations between the neighbouring Wilson lines the volume is large enough so that it can be approximated by the infinite plane, as in the original MV model. For the largest value of $m$ shown in the plot, the resulting gluon distribution is very well described by the Gaussian shape postulated by the $C_{\textrm{GBW}}$ formula. We also take this as a hint that the volume is large enough so that finite volume effects are not visible and also that the large value of $m$ has suppressed the correlations described by the MV model.

\begin{figure}
    \centering
    \includegraphics[width=0.45\textwidth]{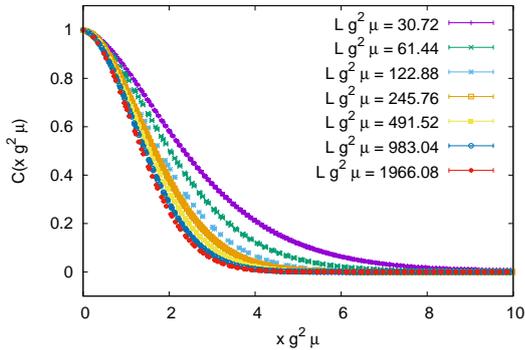}
    \caption{Volume dependence of the dipole amplitude in the MV model on the torus. Increasing torus size pushes the distribution to the left. Consecutive volumes differ by a factor 4 (factor 2 in linear extend). Convergence to some limiting distribution can be seen for very large volumes.}
    \label{fig:MV_volume}
\end{figure}

\begin{figure}
    \centering
    \includegraphics[width=0.45\textwidth]{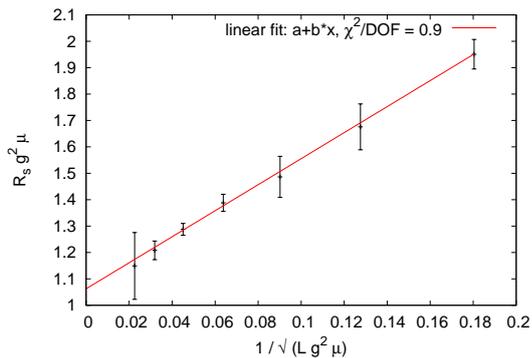}
    \caption{Dependence of the saturation radius extracted from the data from Fig.~\ref{fig:MV_volume} as a function of the fourth root of the volume. Approximate linear scaling can be seen pointing to a very mild suppression of finite volume effects. }
    \label{fig:saturation_radius_volume_dependence}
\end{figure}

We finish this Section by describing the results for the MV model at various intermediate volumes and with values of the infrared regulator and correlations small enough to fall into the logarithmic regime of the Green function Eq.~\eqref{eq: massive green function}. Using that data we can infer the speed of the convergence of the gluon distribution to the infinite volume limit. In Fig.~\ref{fig:MV_volume} we show the gluon distribution for volumes differing by a factor 2, ranging from $g^2 \mu L = 30.72$ up to $g^2 \mu L = 1966.08$. We see slow convergence for increasing volumes. In order to quantify that convergence we use condition given by Eq.~\eqref{eq: saturation radius} to define the saturation radius for each volume separately and plot it against the volume. We do it in Fig.~\ref{fig:saturation_radius_volume_dependence}, where on the horizontal axis we show the square root of the linear extend of the torus. We also add a tentative linear fit, which seems to be consistent with the data at hand. Thus, we can conclude that the finite volume effects vanish very slowly, as $\sqrt{L} = V^{\frac{1}{4}}$, and therefore one needs very large lattices in order to correctly reproduce the MV model from the discretized setting. In the next Section we describe a different approach which is needed to numerically simulate the GBW model which turns out to be free of finite volume effects and in the last Section we will demonstrate that the same method can be applied to the MV model, thus allowing calculations using moderate computer resources.

\section{Discretization of the Golec-Biernat-Wusthoff model}

In the GBW model the shape of the correlation function Eq.~\eqref{eq: correlation function} is postulated to be Gaussian, i.e. given by Eq.~\eqref{eq: S_GBW} and parametrized solely by $R$. A numerical recipe to generate any correlation in position space between two Wilson lines was proposed in  Ref.~\cite{Rummukainen:2003ns}. Here, we describe a slightly modified prescription. Note that there is no physical interpretation of the following steps and quantities introduced below. We are interested in a \emph{technical} algorithm to generate elements of the $SU(3)$ group in each site of the torus such that their correlation function in position space takes a desired form. 

We start with the desired correlation function $C(\mathbf{x})$ defined on the torus. Hence, $\mathbf{x}$ is understood as the shortest distance between two sites on the torus. In order to proceed we calculate numerically the Fourier transform of $C$ getting $C(\mathbf{k})$. Note that in the case of the GBW model we need the Fourier transform of a Gaussian which is a Gaussian, and hence not only $\textrm{Im} \ C(\mathbf{k}) = 0$ but also $C(\mathbf{k}) > 0$. Now, we couple the square root of $C(\mathbf{k})$ to a white noise in the $SU(3)$ algebra. For each momentum $\mathbf{k}$ we generate random variables $\sigma^a(\mathbf{k})$ that are normally distributed with a zero mean and a variance scaled by $c \sqrt{L}$,
\begin{equation}
\langle \sigma^a(\mathbf{k})  \sigma^b(\mathbf{k'}) \rangle =  c^2 L \ \delta^{ab} \delta(\mathbf{k} - \mathbf{k'})  \,.
\label{eq: sigma}
\end{equation}
where $c$ is a technical parameter which controls the ergodicity of the random walk on the $SU(3)$ algebra. After some numerical experiments we set it to a constant value $c=4$. The rescaling of the color sources by the desired correlation function in momentum space
\begin{equation}
    \sigma^a(\mathbf{k}) \rightarrow \chi^a(\mathbf{k}) = \sqrt{C(\mathbf{k})} \ \sigma^a(\mathbf{k})
\end{equation}
guarantees that we obtain in position space the correlation given by $C(\mathbf{x})$. The last missing step is the construction of Wilson lines which are in the $SU(3)$ group. Hence, after a Fourier transform back to position space we exponentiate the so obtained elements of the algebra,
\begin{equation}
    U(\mathbf{x}) = \exp\left( -i \chi(\mathbf{x}) \right)
    = \exp\left( -i \chi^{a}(\mathbf{x}) \lambda^{a}\right)\, ,
    \label{eq: elementary wilson line}
\end{equation}
In order to increase the ergodicity we found that the final Wilson line has to be obtained from a product of a large number of elementary Wilson lines given by Eq.~\eqref{eq: elementary wilson line}. These should not be interpreted as multiple gluon rescatterings, as the above procedure is purely algorithmical and the particular steps do not correspond to any physical process. In our calculations we use 500 elementary Wilson lines.

\begin{figure}
    \centering
    \includegraphics[width=0.45\textwidth]{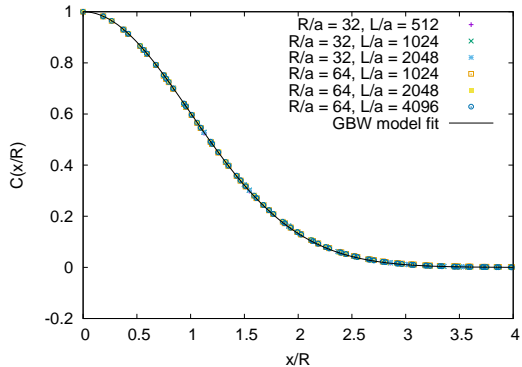}
    \caption{Volume and lattice spacing dependence of the GBW dipole amplitude. Results from different volumes and lattice spacings fall on top of each other providing evidence of negligible finite volume and finite lattice spacing effects.}
    \label{fig:Gaussian_initial_condition}
\end{figure}

We provide results of this procedure in Fig.~\ref{fig:Gaussian_initial_condition} where we plot the unintegrated gluon distribution estimated with various lattice spacings and different volumes. We see clear evidence that at the statistical precision at which we work the procedure is free of finite lattice spacing and finite volume effects.

\section{Discretized McLerran-Venugopalan model revisited}

\begin{figure}
    \centering
    \includegraphics[width=0.45\textwidth]{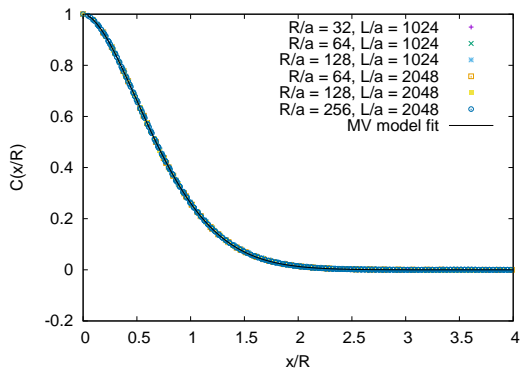}
    \caption{Volume and lattice spacing dependence of the MV dipole amplitude. Results from different volumes and lattice spacings fall on top of each other providing evidence of negligible finite volume and finite lattice spacing effects. $\kappa$ was set to 1.0 in this example.}
    \label{fig:MV_initial_condition}
\end{figure}

It was found that from several phenomenological perspectives the MV model provides a better model for the unintegrated gluon distribution than the GBW model. In particular the behaviour at very small distances, reachable in perturbation theory, has large impact on quantities which can be compared with experimental data. In this region the differences between the MV and GBW models are significant. It would be therefore desirable to dispose of the dipole amplitude given by Eq.~\eqref{eq: S_MV} with a similar quality as shown in Fig.~\ref{fig:Gaussian_initial_condition} for the GBW model. We propose to follow the algorithm presented in the previous Section also for the correlation function Eq.~\eqref{eq: S_MV}. The Fourier transform is real and positive \cite{Giraud:2005vx,Giraud:2016lgg} and therefore it is possible to directly apply Eq.~\eqref{eq: sigma}. All steps of the recipe can be safely performed leading to satisfying results shown in Fig.~\ref{fig:MV_initial_condition}.

\section{Conclusions}

In summary, in this work we have revisited the numerical construction of unintegrated gluon distribution from the McLerran-Venugopalan model. We reminded the results from the model defined on an infinite plane. We then described the differences induced by a compact manifold, for instance two-dimensional torus, which is used typically in numerical calculations. We observed large finite volume effects present in the saturation radius extracted from dipole amplitudes. We identified the origin of these finite volume effects to be the Poisson equation, which is ill-defined for a non-zero net charge on a compact manifold. We then proposed different algorithm to generated the desired dipole amplitude which does not suffer of finite volume and finite lattice spacing corrections. We provided numerical evidence for the correctness of our proposal.

The contribution presented in this work provides a stable numerical setup for studying the effects of the evolution equations. In particular, in view of the extension of the JIMWLK equation by the collinear improvement proposed in Ref.~\cite{Hatta:2016ujq} it is crucial to have full control over the initial condition. The presented setup allows to study the effects of the collinear improvement alone, without any remnants of the finite volume induced by the initial condition. We will present such results in a separate publication.

\section*{Acknowledgements}
The Author thanks K. Cichy, K. Kutak and L. Motyka for many useful discussions. Computer time allocations 'nspt', 'pionda', 'tmdlangevin', 'plgtmdlangevin2' and 'plgquasi' on the Prometheus supercomputer and trial access to the Ares supercomputer, both hosted by AGH Cyfronet in Krak\'{o}w, Poland were used through the polish PLGRID consortium. The Author acknowledges partial support by the Polish National Science Center, grant No. 2017/27/B/ ST2/02755. 

\bibliographystyle{spphys}
\bibliography{references2}

\end{document}